\documentclass[aps,prl,reprint,amsmath,amssymb,superscriptaddress,floatfix]{revtex4-2}
\usepackage{graphicx}
\usepackage[colorlinks, citecolor=blue, linkcolor=blue]{hyperref}

\usepackage{graphicx}
\usepackage{xcolor}

\newcommand{\IPCMS}{Universit{\'e} de Strasbourg, CNRS, Institut de Physique et Chimie des Mat{\'e}riaux de Strasbourg, UMR 7504, F-67000 Strasbourg, France}
\newcommand{\FNRS}{Fonds de la Recherche Scientifique (FRS - FNRS), B-1000 Brussels, Belgium}
\newcommand{\NANOMAT}{Nanomat/Q-mat/CESAM, Universit{\'e} de Li{\`e}ge, B-4000 Sart Tilman, Belgium}
\newcommand{\MAINZ}{Institute of Physics, Johannes Gutenberg University Mainz, 55099 Mainz, Germany}
\newcommand{\juelich}{Peter Gr\"unberg Institut and Institute for Advanced Simulation, Forschungszentrum J\"ulich and JARA, D-52425 J\"ulich, Germany}

\begin{document}

%
%
\title{Dzyaloshinskii-Moriya interaction induced by an ultrashort electromagnetic pulse: \\ Application to coherent (anti)ferromagnetic skyrmion nucleation}

\author{L. Desplat}
\email{louise.desplat@ipcms.unistra.fr}
\affiliation{\IPCMS}

\author{S. Meyer}
\affiliation{\NANOMAT}

\author{J. Bouaziz}
\affiliation{\juelich}

\author{P. M. Buhl}
\affiliation{\MAINZ}

\author{S. Lounis}
\affiliation{\juelich}

\author{B. Dup{\'e}}
\affiliation{\FNRS}
\affiliation{\NANOMAT}
\affiliation{\IPCMS}

\author{P.-A. Hervieux}
\email{paul-antoine.hervieux@ipcms.unistra.fr}
\affiliation{\IPCMS}

\date{\today}

%
%
\begin{abstract}
We show how a Dzyaloshinskii-Moriya interaction can be generated in an ultrathin metal film from the induced internal electric field created by an ultrashort electromagnetic pulse. This interaction does not require structural inversion-symmetry breaking, and its amplitude can be tuned depending on the amplitude of the field.
We perform first-principles calculations to estimate the strength of the field-induced magnetoelectric coupling for ferromagnetic Fe, Co, and Ni, and antiferromagnetic Mn, as well as FePt alloys. Last, using atomistic simulations, we demonstrate how an isolated antiferromagnetic skyrmion can be coherently nucleated from the collinear background by an ultrashort pulse in electric field on a 100-fs timescale.  
\end{abstract}

\maketitle

%
%

%

The study of the ultrafast optical manipulation of magnetism constitutes a fascinating and very active research topic \cite{Siegrist2019}. A breakthrough in this domain could provide the ingredients for the elaboration of a new generation of spintronic devices and lead to computers an order of magnitude faster.
A remarkable result, first observed by Beaurepaire and Bigot in Nickel (Ni) thin films \cite{Beaurepaire1996}, is the ultrafast loss of magnetization occurring within the first 100 fs following a laser pulse. More than 20 years after its discovery, there is no general consensus about the underlying mechanisms of this ultrafast demagnetization, which was attributed to various mechanisms, including, more recently, the spin-orbit interaction \cite{Krieger2015,Stamenova2016}, or the superdiffusive electron transport induced by the laser field \cite{Battiato2010}. This phenomenon is usually referred to as ultrafast thermal demagnetization.   

Another mechanism, not thermally induced, and suggested by Bigot \textit{et al.}~\cite{Bigot2009} is a laser-induced coherent magnetization dynamics--also in Ni thin films, which could be related to a direct coupling between photons and spins, i.e., between the external electric field of the laser and the spins of the electrons in the material. So far, this hypothesis has not been confirmed nor refuted, and is the subject of intense research efforts in the field of ultrafast magneto-optical spectroscopy \cite{Kirilyuk2010,Siegrist2019}.

In Condensed Matter, the interplay between a static external electric field and the magnetization is usually called the magnetoelectic (ME) effect~\cite{Nozaki2019}. It allows the manipulation of the magnetocrystalline anisotropy~\cite{Dieny2017} or that of the Dzyaloshinskii-Moryia interaction (DMI)~\cite{dzyaloshinskii1958a,moriya1960anisotropic,Srivastava2018} with minimal energy consumption, but Coulomb screening limits this effect to a few monolayers~\cite{PhysRevLett.101.137201}.

The situation for time-dependent external electric fields is quite different. In the terahertz (THz) spectral region, the skin depth can, for instance, be as large as a few tens of nanometers \cite{Cuadrado2012}. Moreover, it is nowadays possible to produce powerful ultrashort sub-cycle THz pulses, leading to electric field amplitudes above $10^{11}$~V.m$^{-1}$~\cite{Koulouklidis2020}. Under such conditions, an intense electric field can penetrate inside the material, which may break the inversion symmetry and induce a DMI, resulting in the presence of chiral magnetic states such as skyrmions.
%

Within the search for novel spintronics devices, skyrmion-based designs appear very promising, e.g., as racetrack memories and logic gates~\cite{fert2013skyrmions,sampaio2013nucleation}, for reservoir computing~\cite{prychynenko2018magnetic}, or as reshufflers for probabilistic computing~\cite{pinna2018skyrmion}. Magnetic skyrmions~\cite{bogdanov1989thermodynamically,bogdanov1994thermodynamically} are two-dimensional, non-collinear solitonic spin textures with a nontrivial topology. In ultrathin films and multilayers, small skyrmions are stabilized by an interfacial form of the DMI, which arises under a combination of strong spin-orbit coupling (SOC) and inversion-symmetry breaking at the surface or interface~\cite{fert1990magnetic,crepieux1998dzyaloshinsky,heinze2011spontaneous}.
In particular, antiferromagnetic (AFM) skyrmions have been recently observed in a synthetic antiferromagnet \cite{legrand2020room} and possess attractive properties for applications, i.e., fast internal modes in the THz range, vanishing dipolar fields, and immunity to the skyrmion Hall effect~\cite{zhang2016antiferromagnetic}. As many antiferromagnets are insulating, the optical manipulation of the magnetization in these materials is even more compelling. Recently, the optical trapping of skyrmions was demonstrated in spin-driven chiral multiferroics, where the coupling of an external electric field to the ferroelectric polarization due to the large intrinsic ME effect creates a DMI-like term, which allows the manipulation of skyrmions~\cite{wang2020optical}. The same effect was exploited to coherently switch the polarity and chirality in a magnetic vortex by applying ultrashort electric field pulses~\cite{yu2020nondestructive}.


In this Letter, we show that coherent magnetization dynamics can be induced by a magnetoelectric interaction \textit{created by an ultrashort electromagnetic pulse.}
This mechanism originates from the spin-orbit coupling between the electric field of the pulse and the spins of the delocalized electrons, and reduces to a Rashba SOC. By describing the sample as an ensemble of localized atomic spins embedded in a two-dimensional electron gas (2DEG), and using the RKKY model of indirect exchange, we express the magnetoelectric coupling as a DMI-like term. 
Based on this model, we then show from density functional theory (DFT) calculations that an external electric field is sufficient to create a significant DMI without the need for a structural breaking of the inversion symmetry. Last, we explore the creation of antiferromagnetic skyrmions in metallic thin films, and demonstrate through atomistic simulations that coherent nucleation is possible at the 100-fs timescale, i.e., an order of magnitude faster than thermally-driven skyrmion nucleation processes \cite{Je2018,Berruto2018,Buttner2020}.

%

In the following, we present a simple model to show that the interaction of an external, time-dependent electric field with a magnetic thin metal film leads to the creation of a magnetoelectric effect. 
A very simplified description of the electronic properties of the material sample is based on the distinction between itinerant magnetism carried by the conduction electrons, and localized magnetism carried by the fixed ions. The $s$ electrons are assumed to be at the origin of itinerant magnetism, whereas the $p-d$ electrons are localized around their nuclei to form ionic spins that are responsible for localized magnetism. We assume that the delocalized electrons can be modeled by a 2DEG (ultrathin film geometry), and the light-matter interaction is described by a semi-relativistic expansion of the Dirac-Maxwell mean-field model ~\cite{hinschberger2016coherent}.
As detailed in the Supplemental Material (SM)~\cite{sm}, and in Ref.~\onlinecite{hinschberger2016coherent}, at the lowest order in powers of $1/c$, the spin-light interaction Hamiltonian reads,
\begin{equation}\label{eq:H_SOC}
    \mathcal{H}_{\mathrm{SOC}}=\frac{e\hbar}{4m^2c^2} \mbox{\boldmath$\sigma$} \cdot (\mathbf{E}_{\mathrm{ext}}+\mathbf{E}_{\mathrm{int}}) \wedge \mathbf{p},
\end{equation}
where $m$ is the effective electron mass, $\mathbf{p}$ is the momentum of the electron, $c$ is the speed of light in vacuum, $\mbox{\boldmath$\sigma$}$ is the vector of the Pauli matrices, $\mathbf{E}_{\mathrm{ext}}$ the external electric field of the electromagnetic pulse, and $\mathbf{E}_{\mathrm{int}}$ is the internal electric field solution to the Poisson equation. Thus, the many-electron problem reduces to a one-electron problem, in which the electron experiences SOC with the electric field within the material. 
In the limit of a weak external electromagnetic field--used to demonstrate the linearity of the magnetoelectric effect at low field, the internal field $\mathbf{E}_{\mathrm{int}}$ is proportional to the external field $\mathbf{E}_{\mathrm{ext}}$, therefore $\mathbf{E}_{\mathrm{int}}+\mathbf{E}_{\mathrm{ext}} \propto \mathbf{E}_{\mathrm{ext}}$. The spin-orbit Hamiltonian $\mathcal{H}_{\mathrm{SOC}}$ may be rewritten as a Rashba SOC proportional to $ \left(\mathbf{k} \wedge \mathbf{E}_{\mathrm{ext}} \right)\cdot \mbox{\boldmath$\sigma$}$,
where $\mathbf{k}$ is the electron wave vector.
The effect of the Rashba SOC on a 2DEG has been studied in Refs.~\onlinecite{imamura2004twisted,bouaziz2017chiral}. When a pair of atomic spins $\mathbf{S}_{\mathrm{i}}$ and $\mathbf{S}_{\mathrm{j}}$ are separated by a distance $r_{\mathrm{ij}}$, the field-induced DMI reads
\begin{equation}
| \mathbf{D}^{\mathrm{E}}_{\mathrm{ij}}| = \alpha_{\mathrm{ME}} E_{\mathrm{ext}},
\end{equation}
where $\alpha_{\mathrm{ME}} = \frac{e\hbar}{4m^2c^2} \mathcal{F} (k_{\mathrm{F}}  , r_{\mathrm{ij}})$ is the antisymmetric magnetoelectric coupling, in which $k_{\mathrm{F}}$ is the Fermi wave vector, and $\mathcal{F}$ is defined in the SM~\cite{sm}. In particular, for an external electric field applied along $z$, the DMI vector has the same symmetry as that of interfacial DMI. Let us stress that in the absence of an electric field, $| \mathbf{D}^{\mathrm{E}}_{\mathrm{ij}} | = 0$.

%

Next, we perform density functional theory calculations in order to estimate the strength of the DMI created by an external electric field. To this extent, we apply the full-potential linearized augmented planewave (FLAPW) method, as implemented in the FLEUR code \cite{FLEUR}. 
We consider model systems in the form of 3\textit{d} transition metal unsupported monolayers (UMLs), in which the \textit{d}-band is progressively filled from Mn to Fe, Co, and Ni, as well as unsupported trilayers of Fe/Pt/Fe, Pt/Fe/Pt, and Pt/Mn/Pt. We use the lattice parameters and crystal structures of bulk Mn, Fe, Co, Ni, and L$_{10}$ PtFe or PtMn binary alloys \cite{Alsaad2020}. A static electric field is applied perpendicular to the film plane, which allows us to circumvent the calculation of the long-range Coulomb-type interaction present in bulk systems. This drastic approximation thus only holds at ultrafast timescales, at which the mobility of the electrons is mainly influenced by their effective mass, rather than by scattering events. 

\begin{figure}
\includegraphics[scale=0.9]{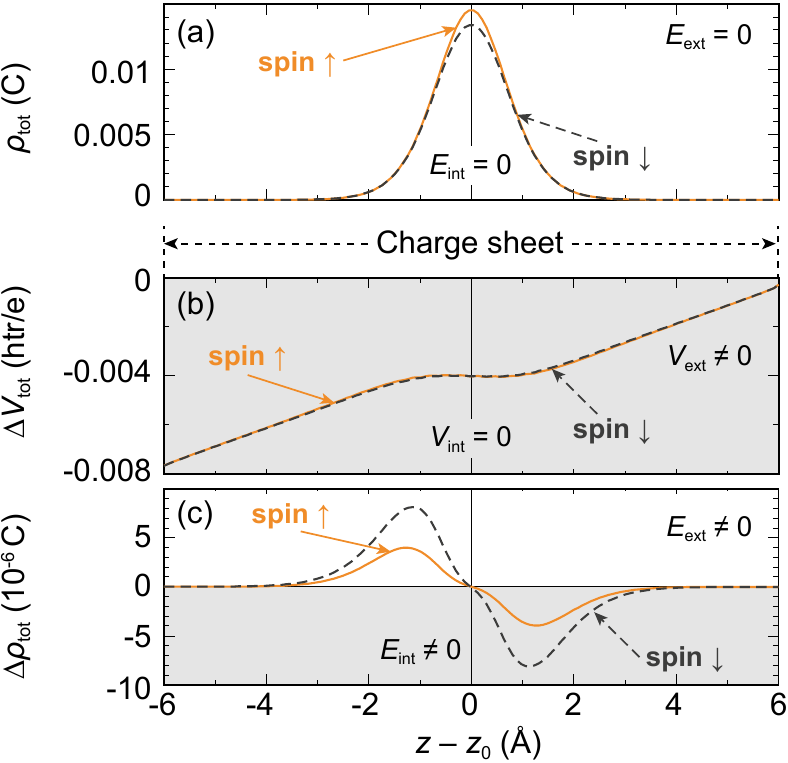}
\caption{Charge density and potential in an Fe unsupported monolayer (UML) along the $z$-direction perpendicular to the film. (a) Charge density for spin up ($\uparrow$) and spin down ($\downarrow$) without applying an external electric field. $z_0$ denotes the position of the atomic center whereas the $y$ position lies in the middle of the unit cell. (b) Potential difference for both spins under an electric field of $E_\textrm{ext} = +6.66\times10^8$V.m$^{-1}$ with charged sheets at position $\pm 6\,\textrm{\AA}$. (c) Difference of charge density after applying the electric field.
}
\label{fig:potential and charge density}
\end{figure}
The calculations can be understood based on the example of a Fe UML, as shown in Fig.~\ref{fig:potential and charge density}. In the absence of an external electric field (Fig.~\ref{fig:potential and charge density}(a)), the system is centrosymmetric, so the charge density for both spin up ($\uparrow$, orange line) and spin down ($\downarrow$, dashed black line) as a function of the vertical distance from the center of an atom is symmetric, and no internal field is created  ($E_\textrm{int} = 0$). In that case, there is no DMI. This behavior holds for all centrosymmetric, unsupported monolayers.

We then look at the effect of an external electric field, which can be applied by adding two oppositely charged sheets below and above the film. Here, positive fields are related to a positively charged sheet below, and a negatively charged sheet above the film.
The potential difference created by an applied field of $6.66\times10^8$~V.m$^{-1}$ is shown in Fig.~\ref{fig:potential and charge density}(b), for which the charged sheets are placed at $\pm$6~\AA\ from the film surface. The sheets create a linear potential along $z$, which vanishes in the film region due to Coulomb screening.
The effect of the latter is visible in the difference in the charge density compared to the zero-field scenario, as shown in Fig.~\ref{fig:potential and charge density}(c). In the presence of the external field, the mirror symmetry of the charge density is broken, and the charges recombine depending on their spin channels. The charge density differences exhibit two peaks: A positive one for $z<0$, and a negative one for $z>0$, which respectively translates an accumulation or a depletion of charges in the film. This charge imbalance creates an internal electric field $E_{\mathrm{int}}$, which can act as a spin-orbit coupling contribution.

\begin{figure}
\includegraphics[scale=1]{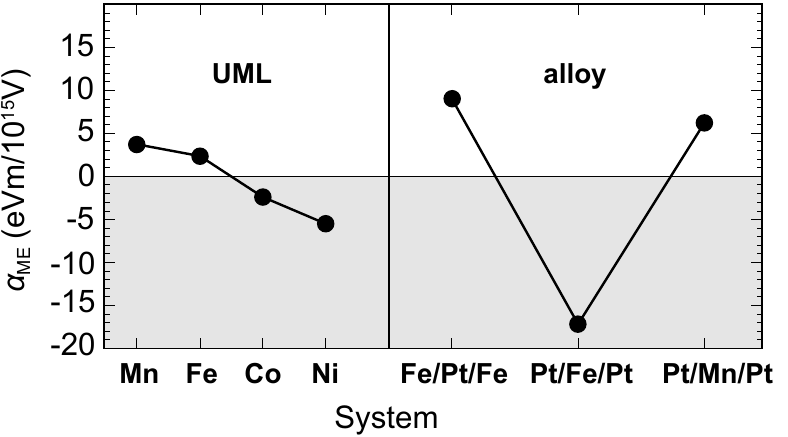}
\caption{DFT-calculated values of the magnetoelectric coupling, $\alpha_\textrm{ME}$. Positive (negative) values correspond to increasing strength of clockwise- (counterclockwise-) preferring DMI with the applied electric field.}
\label{fig:Electric field results DFT}
\end{figure}

To confirm this hypothesis, we calculate the SOC contribution in the presence of such an internal electric field. To do so, we perform self-consistent spin-spiral calculations \cite{Sandratskii1991}, where SOC is added to first order in perturbation theory \cite{Heide2009}.
The strength of the field-induced SOC, which we interpret as a DMI term $D_\textrm{eff}^E$, is then determined in the limit of $q \rightarrow 0$ in the vicinity of the magnetic ground-state, as a function of the external electric field.  As expected, the DMI depends linearly on the electric field~\cite{sm}, which allows the extraction of the linear ME coefficient, $\alpha_{\mathrm{ME}}$. Figure~\ref{fig:Electric field results DFT} shows the value of $\alpha_{\mathrm{ME}}$ for different $d$-band fillings of the 3$d$ transition-metal monolayers.
For Mn and Fe, the DMI favors a right-rotating spin spiral, and $\alpha_{\mathrm{ME}}$ goes from $4\times10^{-15}$ eV.m.V$^{-1}$ for Mn, to $2\times 10^{-15}$ eV.m.V$^{-1}$ for Fe. As the $d$-band filling increases, i.e., for Co and Ni, the DMI changes sign, and $\alpha_{\mathrm{ME}}$ reaches $-4\times 10^{-15}$ eV.m.V$^{-1}$ for Ni. 

To increase the value of $\alpha_{\mathrm{ME}}$, we associate 5$d$ elements to the 3$d$ ultrathin films. In bulk, these combinations could be achieved by exploring binary alloys such as FePt, CoPt or MnPt. In this case, $\alpha_{\mathrm{ME}}$ can reach values up to $-17\times 10^{-15}$ eV.m.V$^{-1}$ for Pt/Fe/Pt. This corresponds to a DMI of 
17 meV/magnetic atom for an electric field amplitude of $10^{12}$ V.m$^{-1}$, which, as we show in the following, is sufficient to nucleate isolated skyrmions at the 100-fs timescale. 
%
%

 %

Finally, we perform atomistic simulations to demonstrate how this new mechanism can be used to nucleate an AFM skyrmion coherently, by applying an ultrafast electromagnetic pulse. To that extent, we simulate an ultrathin magnetic film consisting of $N$ magnetic moments $\{ \mathbf{\hat{m}}_{\mathrm{i}} \}$, $i=1 \hdots N$, of norm unity, on a two-dimensional simple square lattice.  The grid size is $50 \times 50$ with periodic boundary conditions at the edges. The pulse is assumed to be linearly polarized, and in an oblique incidence configuration, which produces a longitudinal component $E_z$ inside the film.  We give the effective Heisenberg Hamiltonian,
 \begin{equation}
    \begin{split}
        & \mathcal{H} =   - J_{\mathrm{eff}} \sum_{\langle ij \rangle}  \mathbf{\hat{m}}_{\mathrm{i}}  \cdot \mathbf{\hat{m}}_{\mathrm{j}}
    - K \sum_{\mathrm{i}} m_{z,\mathrm{i}}^2 \\
    & - \sum_{\langle ij \rangle} \left(D_{\mathrm{eff}}^{\mathrm{interf}} + \alpha_{\mathrm{ME}} E_{z,\mathrm{i}}\right) \left( \mathbf{\hat{z}} \times \mathbf{\hat{r}}_{\mathrm{ij}}\right) \cdot \left(  \mathbf{\hat{m}}_{\mathrm{i}} \times \mathbf{\hat{m}}_{\mathrm{j}}  \right),
    \end{split} 
\end{equation}
where the double summations are carried out over all pairs of first nearest neighbors $\langle \mathrm{ij} \rangle$, $J_{\mathrm{eff}}<0$ is the effective AFM Heisenberg exchange coupling constant, $K$ is the perpendicular magnetic anisotropy constant, $D_{\mathrm{eff}}^{\mathrm{interf}}$ is the effective interfacial DMI coupling at zero electric field, $E_{z,\mathrm{i}}$ is the perpendicular electric field at site i such that $D_{\mathrm{eff},\mathrm{i}}^E = \alpha_{\mathrm{ME}} E_{z,\mathrm{i}}$,  and $\mathbf{\hat{r}}_{\mathrm{ij}}$ is the unit displacement vector between first neighbors. 
We furthermore define reduced parameters as $d=D_{\mathrm{eff}}^{\mathrm{interf}}/| J_{\mathrm{eff}}|$, $k=K/|J_{\mathrm{eff}}|$, and $d_E = \alpha_{\mathrm{ME}} E_{z}/| J_{\mathrm{eff}}| $. Simulations are performed at zero temperature by solving the Laudau-Lifshitz-Gilbert equation~\cite{brown1963thermal} with a Gilbert damping of $\lambda=0.3$, and $J_{\mathrm{eff}}=-11$~meV. The magnetization is initialized in the AFM state with a single in-plane defect in the center, such that an electric field exerts a nonzero torque on the magnetic moments. At $t=0$, the  electric field is applied in the lattice center, with a spot diameter of ten lattice sites, and a time dependence taken as a Heaviside step function. While electromagnetic sources such as lasers typically produce mm-$\mu$m-wide spots~\cite{manfredi2005nonlinear}, here, no new physics would emerge by simulating a wider spot since the DMI exerts no torque on the collinear state at zero temperature.
Last, the total topological charge $Q$ of the vector of the AFM order, $\mathbf{L}=(\mathbf{M}_1 -\mathbf{M}_2) /2 $, where $\mathbf{M}_{1,2}$ correspond to the AFM sublattices, is computed using a discretized description~\cite{berg1981definition,bottcher2018b} and used to track the skyrmion nucleation.
\begin{figure}
\includegraphics[width=1\linewidth]{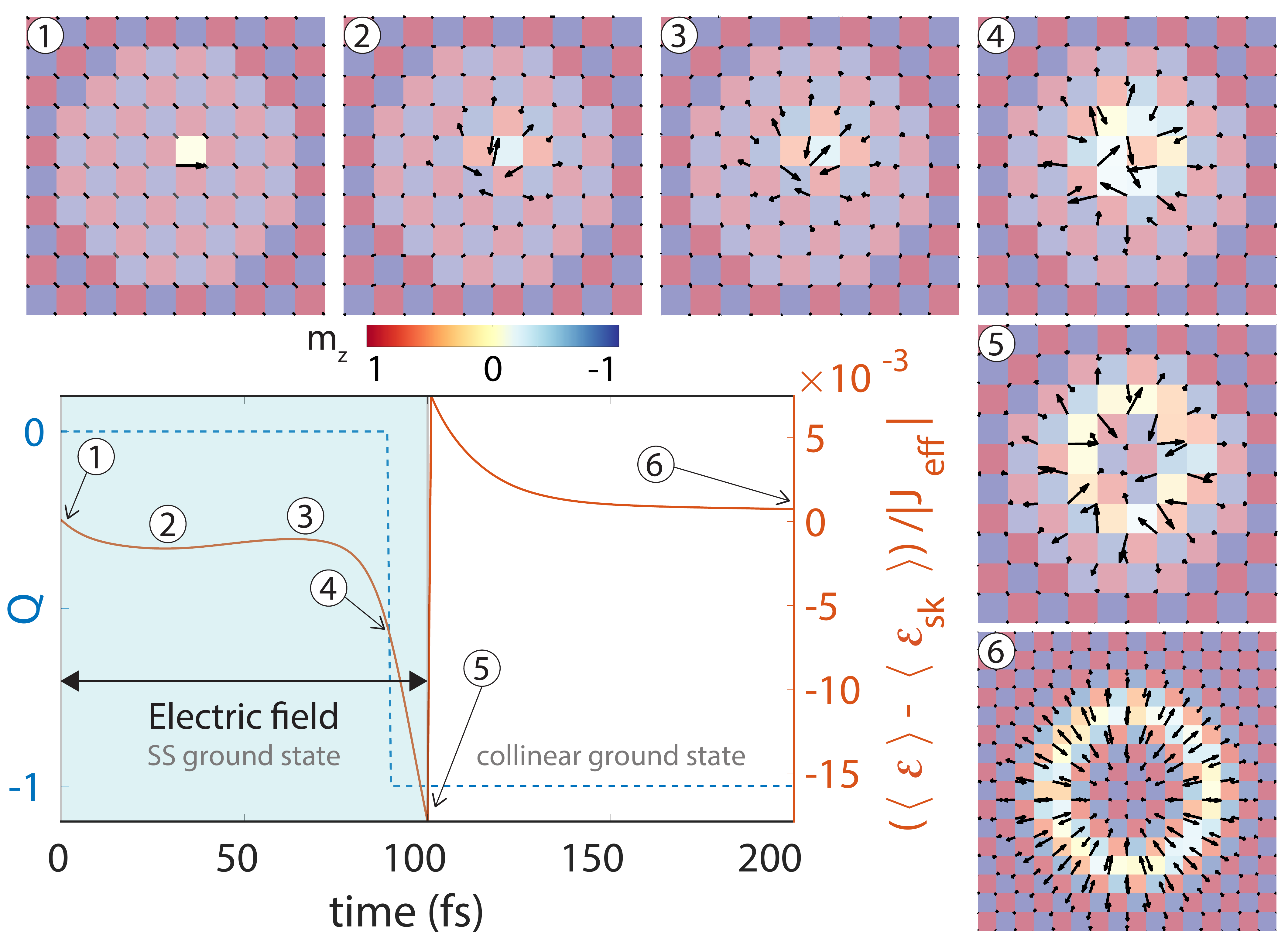}
\caption{AFM skyrmion creation under a pulse in electric field. The graph shows the time evolution of the topological charge (left, dotted line) and of the mean energy per site (right, solid line). The blue shaded area indicates the duration of the electric field. The spin maps show snapshots of the magnetization around the lattice center at different times of the simulation. The beam spot is shown as an overlayed white disk. The parameters of the simulations are $k=0.5$, $d=0.44$, $d_E=1.13$.
}
\label{fig:E_Q_vs_t}
\end{figure}

The nucleation mechanism is illustrated in Fig.~\ref{fig:E_Q_vs_t}. We show the time evolution of the total topological charge of the system, $Q$ (left), and of the mean energy per site normalized by the exchange coupling constant, $\big( \langle \mathcal{E} \rangle - \langle \mathcal{E}_{\mathrm{sk}} \rangle \big) /| J_{\mathrm{eff}}|$  (right), where $\langle \mathcal{E}_{\mathrm{sk}} \rangle$ is the mean energy per site of the partially relaxed skyrmion at $t = 1000$~fs. The blue shaded area indicates the duration of the electric field, which is 100 fs in this case. The spin maps correspond to snapshots of the magnetization around the center of the simulated area at different times. When the field is switched on ($t=0$), the mean energy first decreases towards a local minimum corresponding to a perturbed collinear state (snapshot 2, 29 fs) and increases again to a maximum--the barrier top for the skyrmion nucleation in this scenario (snapshot 3, 64 fs), before dropping brutally as a unit of topological charge is nucleated (snapshot 4, 90 fs). This behavior stems from the fact that the increase in the DMI induced by the field momentarily changes the ground state of the system in the beam cross-section, from the collinear to the spin spiral (SS) state. It follows that the skyrmion state becomes more favorable that the collinear state and is more easily nucleated during the relaxation process.  Snapshot 5 corresponds to the skyrmion state at the time when the beam is turned off (100 fs). This results in a discontinuous jump in energy, at which point the collinear ground state is restored, followed by the relaxation of the newly formed metastable skyrmion (snapshot 6). We note that we used the minimum value of $d_E$ that enables nucleation, and that, for a larger value, the nucleation process is typically even faster.
%

\begin{figure}
\includegraphics[width=.9\linewidth]{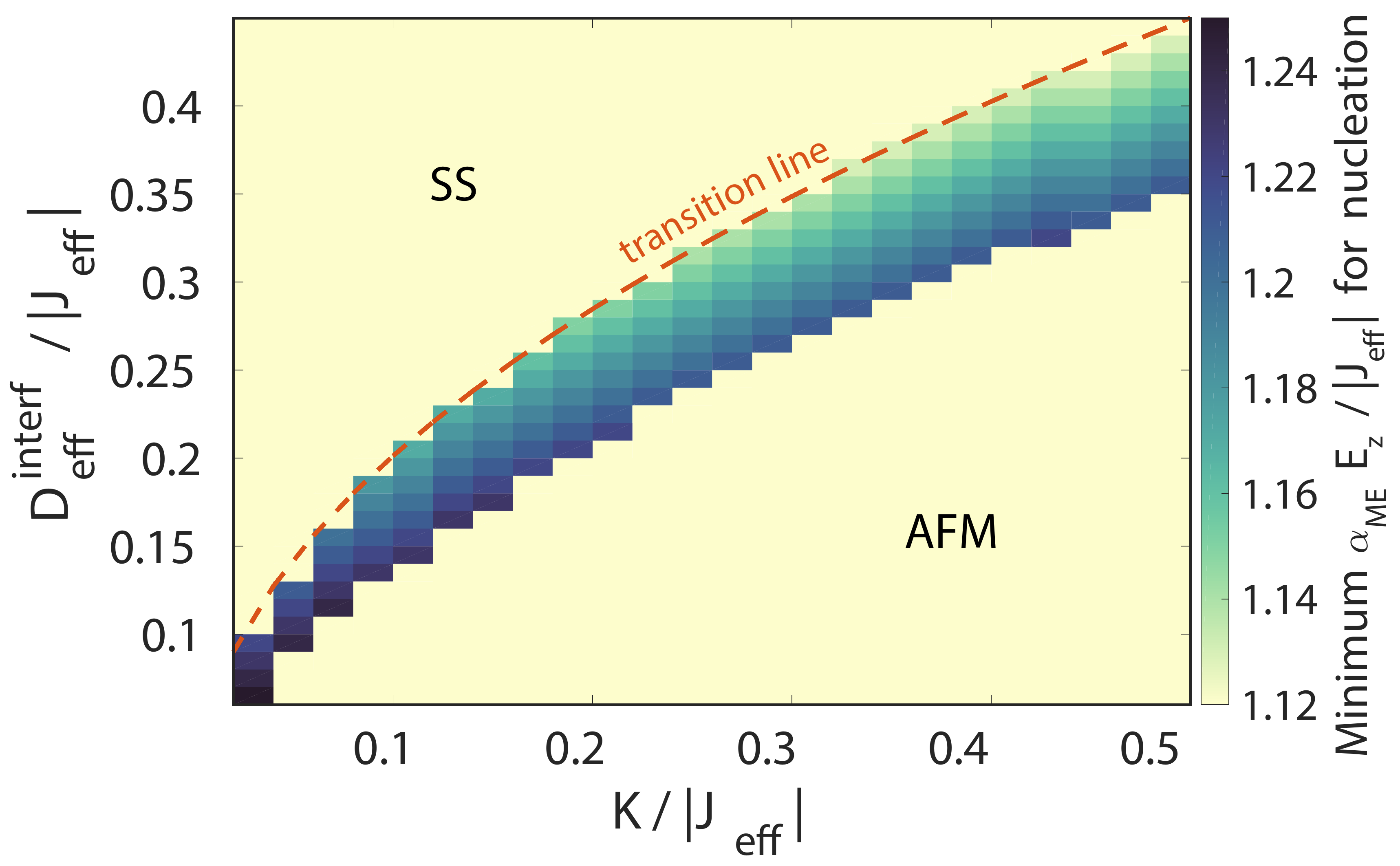}
\caption{Minimum DMI induced by an electric field required for the nucleation of a metastable AFM skyrmion. The area with the color gradient corresponds to the region of existence of metastable skyrmions as a function of anisotropy and effective interfacial DMI at zero electric field. 
The colorscale indicates the minimum value of the field-induced DMI required for skyrmion nucleation for $J_{\mathrm{eff}}=-11$~meV and a damping of $\lambda=0.3$. }
\label{fig:phase_diag}
\end{figure}

In order for a skyrmion to survive at zero electric field, the material parameters must allow skyrmions to be metastable. In Fig.~\ref{fig:phase_diag}, the area with a color gradient indicates the existence of metastable skyrmion excitations on the collinear background as a function of reduced DMI and anisotropy~\cite{wilson2014chiral,leonov2016properties,bessarab2019stabilityafmsk}. 
The red dotted line marks the transition line between the spin spiral ground state and the collinear ground state as $d_c=2\pi^{-1} \sqrt{k}$~\cite{bogdanov2002magnetic,heo2016switching}. The colorscale indicates the minimum value of the reduced DMI induced by the electric field, $d_E$, that is required to nucleate a skyrmion for each set of $(d,k)$. We find the smallest value of $d_E=1.12$ in the large $(d,k)$ region, which also corresponds to the fastest topological charge nucleation in 90~fs, and a maximum value of $d_E=1.25$ at low $(k,d)$, at which the topological charge takes a maximum of 210~fs to nucleate. With our choice of parameters and the value of $\alpha_{\mathrm{ME}}$ for Pt/Mn/Pt, we obtain a minimum applied field of $E_z=2\times 10^{12}$ V.m$^{-1}$. In the SM, we show similar results for the case of a ferromagnetic skyrmion nucleation~\cite{sm}, for which the minimum $d_E$ is 1.52, yielding a minimum field $E_z=10^{12}$ V.m$^{-1}$ for Pt/Fe/Pt. In both cases, the required field values are realistically achievable~\cite{Koulouklidis2020}.

%

In conclusion, we have shown that at ultrashort timescales, an external electric field can induce an internal electric field in metals, which breaks the inversion symmetry and creates a DMI. We have shown how this mechanism allows a coherent nucleation of skyrmions on a 100-fs timescale. This result provides a new handle for the ultrafast manipulation of the magnetization, and opens the way to an optical control of skyrmions, e.g., via ultrafast laser pulses in metallic thin films.

\begin{acknowledgments}
We thank Melanie Dup\'e and Giovanni Manfredi for useful discussions. This work used the ARCHER UK National Supercomputing Service (http://www.archer.ac.uk) and was supported by the University of Strasbourg Institute for Advanced Study (USIAS) via a Fellowship, within the French national program ``Investment for the Future'' (IdEx-Unistra). B. Dup\'e and S. Meyer acknowledge support via DARPA Grant No.
HR0011727183-D18AP00010 (TEE program). B. Dup\'e and P. Buhl acknowledge funding by the DFG under Grant No. DU 1489/3-1.
\end{acknowledgments}

\bibliography{articles}
\end{document}